\documentclass[11pt]{article}
\begin{document}
\begin{flushright}

July 2009\end{flushright}
\vspace{1.7in}
\begin{center}\Large{\bf Has the Born rule been proven?}\\
\vspace{1cm}
\normalsize\ J. Finkelstein\footnote{
        Participating Guest, Lawrence Berkeley National Laboratory\\
        \hspace*{\parindent}\hspace*{.5em}
        Electronic address: JLFINKELSTEIN@lbl.gov}\\
        Department of Physics and Astronomy\\
        San Jos\'{e} State University\\San Jos\'{e}, CA 95192, U.S.A
\end{center}
\begin{abstract}
This note is a somewhat-lighthearted
comment on a recent paper by David Wallace entitled ``
A formal proof of the Born rule from decision-theoretic assumptions''.
\end{abstract}
\newpage
The many-worlds interpretation (MWI) of quantum theory asserts that the    
universe can be represented by a deterministically- and unitarily-evolving
quantum state.  A major challenge to this interpretation has been the 
justification of the values, and even of the meaning, of the probabilities 
which are predicted by the Born rule and which are in such magnificent
agreement with all observations.  Some time ago, David Deutsch [1] suggested
that decision theory could provide this needed justification, by leading to 
what I shall call the decision-theoretic Born Rule, and abbreviate as DTBR.
This rule, (as formulated in [2]), is
\begin{quote}
{\bf Decision-theoretic Born Rule:} A rational agent who knows that the 
Born-rule weight of an outcome is $p$ is rationally compelled to act as if 
that outcome had probability $p$. \end{quote}

In a series of papers, David Wallace [2, and references therein] has elaborated
upon Deutsch's suggestion; I shall refer to the attempt to establish
the DTBR as the D-W program. Two of the recent critics of this program are
Price [3] and Kent [4], where references to earlier critical papers\footnote
{including one [5] by the present author [advertisement]} can be found. 

In this note I comment on the most recent paper by Wallace [2].  After a 
very brief review of the D-W program as pursued in [2], I will describe three 
different possible agents whom I will call  the Egalitarian, the 
Optimist, and the Stoic.  I will argue that each of them should be considered 
to be acting rationally, although none of them will obey the DTBR.  
The first two 
of these agents will violate at least one of the axioms of rationality 
adopted by Wallace in  [2]; of course I am hoping that the 
conclusion the reader will draw from this is not that these agents are, 
after all, irrational, but rather that
Wallace's axioms are too strong.  The third agent will respect {\em all} of
the axioms of ref. [2], but nevertheless will not obey the DTBR.

\vspace{3ex}
\large{\bf The D-W program:} \normalsize  Consider an "agent" who is offered
the choice of several "games".  Each game will consist of a (quantum) event 
with several possible outcomes, after which the agent will receive a "reward" 
depending on the outcome of the event.  I will write $r_i $ for the reward
given to the agent following the ith outcome, and ${\cal V}(r_i)$ for the 
value the agent places on receiving   $r_i$.  For the agents I describe
below, I will take rewards to be numbers of dollars (which could be negative),
 and assume that  ${\cal V}(\$ x)=\$ x$,
but this need not be true in general.

According to the MWI, after a quantum event the state of the world is a
superposition of branches, corresponding to the various possible 
outcomes of the event.  That state can be written
\[
|\Psi \rangle = \sum _{i} c_{i}|r_{i}\rangle . \]
where $|r_{i}\rangle $ represents a branch in which outcome $i$ had 
occurred
and in which the (descendant of the) agent has received reward $r_i$, and where
$\sum _{i}|c_{i}|^{2} = 1$.  The "quantum weight" associated with $r_i$,
which I will write as $w(r_{i})$,
is $|c_{i}|^2$; according to the Born rule, this is the probability that the 
agent will receive $r_i$. For a game G, the quantum expectation of 
$\cal V $ is
\[
\langle {\cal V}\rangle (G) = \sum _{i} w(r_{i}){\cal V}(r_{i}),
\]
where of course the rewards $r_i$  and the weights $w(r_i)$ on the 
right-hand-side of this equation are those appropriate to the game G.  

The agent is assumed to have a preference order for games; 
the D-W program seeks to establish 
the DTBR by showing that the only rational 
preference order is given by, for any games $A$ and $B$, $A$ is
preferred to $B$  exactly 
when  $\langle {\cal V} \rangle (A)> \langle {\cal V}\rangle (B) $.
To accomplish this, Wallace in ref.\ [2] adopts several axioms which he
argues any rational preference order must obey.  One of these is, in 
paraphrase,
\begin{quote}
{\bf Diachronic Consistency:} Suppose that an agent plays a game $G$, 
and that, for each $i$, his descendant  in branch $i$ has a choice of game
$H_i$ or $H_{i}^\prime $; then

i) If none of the descendants prefer $H_{i}^\prime $ to $H_i$ , then 
the agent must not prefer playing $G$  followed by $H_{i}^\prime $
to playing $G$ followed by $H_i$.

ii) If in addition at least one  descendant prefers  $H_i$ to
$H_{i}^\prime $, then the agent must prefer playing $G$ followed by
$H_i$ to playing $G$ followed by $H_{i}^\prime $.
\end{quote}

Another axiom which Wallace argues to be required by rationality is
\begin{quote}
{\bf Solution Continuity:} If the agent prefers game $G$ to game 
$G^\prime $, and if game $H$ is sufficiently close to $G$ and 
$H^\prime $ sufficiently close to $G^\prime $, then the agent must
prefer $H$ to $H^\prime $.
\end{quote}

The three agents I will define below do not obey the DTBR. The 
Egalitarian will violate
diachronic consistency, and the Optimist will violate both 
diachronic consistency and 
solution continuity; of course the implication of this is, if these two
agents are judged to be rational, that these axioms are not really
required by rationality.   The Stoic will respect {\em all} the axioms
adopted in [2].

\vspace{3ex}
\begin{large} {\bf The Egalitarian:}\end{large} The
Egalitarian wishes that all of his descendants fare as equally as is
possible.  Unlike the egalitarian described by Greaves [6], this
Egalitarian does not try to apply equal weights to all branches; instead,
at least in the cases in which the $\langle {\cal V}\rangle $ of two 
games are equal, he prefers the game which makes the
differences of the rewards received by 
his descendants smaller.  So, for example, he would prefer game $A$ 
(receive either \$2 or \$3, with equal quantum weight) to game $B$
(receive either \$1 or \$4, with equal quantum weight) even though
$\langle {\cal V}\rangle $ is the same for those two games; thus he 
would violate the DTBR.

Wallace would say that the Egalitarian is irrational, because in 
some cases he would not be diachronically consistent.
Suppose that the Egalitarian has two immediate descendants whom
I will call $D_1$ and $D_2$ (each of whom is an Egalitarian, 
of course) and that these respectively have descendants 
$D_{1}$-junior and $D_2$-junior.  The senior Egalitarian would
want  $D_{1}$-junior   and $D_2$-junior to fare equally; they are,
so to speak, his grandchildren.  But neither $D_1$ nor $D_2$ 
would have that concern; $D_{1}$-junior is not a descendant of
$D_2$, and  $D_2$-junior is not a descendant of $D_1$.  The fact that
the senior Egalitarian has a concern which none of his descendants
share could certainly lead to a violation of diachronic consistency.
But is it irrational?

In fact, if an agent's preferences are to obey the DTBR, he {\em must}
have concerns not shared by any of his descendants; he must be 
concerned with the quantum weights of the ensuing branches, while his
descendants will not be (do you even know the quantum weight of the
branch you  inhabit?).
In real life, it is not remarkable, and not considered irrational, to
have concerns which one knows that ones descendants (or one's
future self) will not share, and in some cases this does lead to a
violation of diachronic consistency. Wallace himself describes an example 
of this, and then  concedes that ``isolated occurrence'' of violation
of diachronic consistency might not be irrational, but he asserts
that ``In the presence of {\em widespread, generic} violation of
diachronic consistency, agency in the Everett universe is not 
possible at all.''  One might disagree with this assertion (as Kent [4]
does), but in any case it does not seem a sufficient reason to 
adopt an axiom which forbids {\em any} violation.  Wallace does
point out that ``Everettian branching is ubiquitous; agents branch 
all the time (trillions of times per second, at least...)'', but it is not
clear how this changes anything.  Surely an agent is not called upon
to make decisions trillions of times per second, so it seems
that it would take more than the requirement of rationality to 
forbid him to make decisions which might occasionally lead to a
violation of diachronic consistency. 

\vspace{3ex}
\begin{large}{\bf  The Optimist:}\end{large}
Consider rewards to be monetary, and set ${\cal V}(\$ x)=\$ x$.
Now define, for a game $A$, $LR(A)$ to be the largest of the rewards 
offered by $A$.  Then the Optimist prefers game $A$ to game $B$
when $LR(A)>LR(B)$. This Optimist is similar to the ``future self elitist''
discussed in [4] except that he does not resolve ties.  

Could this be a rational preference? Suppose,
first, that the Optimist were so frail that he would immediately expire 
if, when playing  a game $A$, he suffered the disappointment of 
not receiving $LR(A)$.  This would mean that all of his surviving descendants
would certainly be in the   $LR(A)$ branch, so why should he care
about the rewards in other branches?  (Note that if he were this frail 
but subscribed to some ``one-world'' interpretation of
quantum theory, his preferences would be quite different; it is the MWI
which guarantees that he will have a descendant in the  $LR(A)$ branch.)
Note also that if  in fact he was not so frail, but merely 
believed that he was,  his preferences would equally seem to be rational.

Admittedly, it is somewhat far-fetched to imagine that a person would be,
or even believe himself to be, frail in just this way.  On the other
hand, there are less far-fetched circumstances in which
the Optimist's preferences would seem rational. So let us allow that
the Optimist will have surviving descendants on several branches,
and suppose that he was convinced by Saunders [7] that
the MWI implies that he should expect to become {\em one} of his 
descendants;
then he might believe that he would be lucky enough to become the one
in the   $LR(A)$ branch.\footnote {Just as in the classical case, the
Optimist must be careful not to fall victim to a Dutch book. He would want
to play (i.e.\ prefer to the null game) a game in which a   coin was tossed
and he would win $\$1$ if the coin showed heads but would lose $\$2$
if the coin showed tails.  Likewise, he   would want to play a game in which 
he would win $\$1$ if tails, but lose $\$2$ if heads.  However, he would 
not want to play both games if they depended on a single coin toss, 
because for the combined game $LR$ is negative} 
It might be callous of him to not be 
concerned about the rewards received by all the other descendants,
but callousness is not usually considered to be irrational.

In some cases, the Optimist's preferences can violate Wallace's axiom of 
diachronic consistency.  Suppose that after playing a game $G$ he had
two descendants called $D_1$ and $D_2$, and that $D_1$
could play either $H_1$ (which rewards \$2 with certainty)
or $H_{1}'$ (which rewards \$1 with certainty),  while $D_2$ could 
play either $H_2$  or $H_{2}'$ (each of which rewards \$3 with 
certainty).  Then $D_1$ would prefer $H_1$  to $H_{1}'$, $D_2$
would be neutral between $H_2$ and $H_{2}'$, but the Optimist
would be diachronically inconsistent because he would not prefer 
$G$ followed by
$H$ to $G$ followed by $H'$.  In this case the Optimist would be
confident that he would become $D_2$, so it does not seem irrational
for him not to care about what   $D_1$ wants.

The preferences of the Optimist also violate Wallace's axiom of 
solution continuity,
as can be seen from the following example:  Suppose the Optimist
is choosing between the following two games:

\vspace{1ex}game $A$ := receive \$1 with certainty, and

game $B_{\epsilon}$ := receive \$1 or \$0, with $w(\$1)= \epsilon$ 
and $w(\$0)=1-\epsilon$.

\vspace{1ex}\noindent
According to the strategy I have defined for him, the Optimist should
prefer $A$ if $\epsilon=0$, but should be indifferent between $A$ and
$B_{\epsilon}$ for any $\epsilon>0$.  However, in justifying his axiom
of solution continuity, Wallace points out ``Any discontinuous preference
order would require an agent to make arbitrarily precise distinctions
between different acts, something which is not physically possible''.
In particular, this means that the Optimist's strategy could not, 
in practice, be carried out, because it is
not physically possible to know with infinite precision what game is being
played.  Let me call this the ``precision limitation'' and agree with 
Wallace that it implies that our Optimist can never know that $\epsilon$
is precisely zero.

One possible response to the precision limitation is to shrug 
``So who cares?''
After all, the title of Wallace's paper is ``A formal proof of the Born 
rule...'', not ``A practical guide to selecting games''.  According to the
MWI, no such guide is needed;    the MWI is deterministic, so arguably
it implies that the agent is not free to make any choices at all, which
would mean that
{\em no} strategy is in practice possible.  Wallace is of course aware of
this problem; he responds to it in a
footnote, where he writes ``...we can talk about rational strategies even
if an individual agent is not free to choose whether or not his strategy
is rational.''  Yes we can, but if we are just talking about strategies, without
requiring that an agent could actually pursue them, it is not clear why
we should limit our talk to strategies which, according to some other,
non-deterministic theory, an agent might actually pursue.

The Stoic, described  below, offers a different response to the precision
limitation; he will {\em exploit} that limitation to define a strategy which
will satisfy all of Wallace's axioms, but which nevertheless will be in
conflict with the DTBR.

\newpage
\begin{large}{\bf  The Stoic:}\end{large}
For the Stoic, I again take ${\cal V}(\$ x)=\$ x$. Also, just like the 
Optimist, the Stoic always expects to receive the greatest 
award offered by any game he plays; so, like the Optimist, if asked
to choose between the two games considered above:

\vspace{1ex}game $A$ := receive \$1 with certainty, and

game $B_{\epsilon}$ := receive \$1 or \$0, with $w(\$1)= \epsilon$ and
$w(\$0)=1-\epsilon$,

\vspace{1ex}\noindent  he will have no preference in the case in 
which he knows that 
$\epsilon >0$.  But the Stoic is also impressed by the fact that the 
precision limitation means that he can never know that $\epsilon=0$.
Suppose that he knows that the value of $\epsilon$ lies within a certain
interval, but he does not know where within that interval it does lie.  
If the point 0 is not included in the interval, that means he is certain
that $\epsilon \neq 0$; if the point 0 is included in the interval, then, 
since 0 is just a single point, he still thinks that, with probability one, 
$\epsilon \neq 0$.  So it is reasonable for him in either case to act as 
if  $\epsilon \neq 0$; that is, he will always be neutral between games
$A$ and $B_\epsilon$.

Another way to understand the Stoic is to imagine that a bookmaker
were to say to him ``You can choose $A$ or $B_0$; if you choose
$A$, I will toss a fair coin and give you \$1 whichever way the 
coin lands; if you choose $B_0$ I will toss a fair coin 
and give you \$0 whichever way
the coin lands.''  The Stoic would rather receive \$1 than \$0.  However,
he thinks ``The bookmaker must have at least \$1 in his pocket, so that
he could give it to me if I were to choose $A$. If I choose $B_0$, he
is not obligated to give me anything.  Nevertheless, there will be some
branch, albeit with minuscule quantum weight, in which the money
which began in the bookmaker's pocket winds up, via quantum 
tunneling, in my pocket.  Therefore if I choose $B_0$, it is possible 
that I will receive at least \$1.''  Then, like the Optimist, he expects that
if he chooses   $B_0$ he {\em will} receive at least \$1.  Since he also
expects that if he chooses $A$ he will receive at least \$1, he is neutral
between $A$ and $B_0$.

We can easily understand what will be the Stoic's preference order in
general, if we imagine that there is a finite award which is the greatest
award that any game can offer.  Write that award $\$max$, and think
of it as ``all the money in the universe''\footnote{  This might seem 
inconsistent,
since if you gain $\$max$ in each of two games you have gained
$2(\$max)$ in the combined game.  This seeming inconsistency is due
to the simplification of letting the utility of a monetary reward equal the
reward, and that need not be true.  If you already have all the money
you could possibly use, it is arguably no better to have twice that 
amount, so it is not inconsistent to assume there is a reward so great
that its utility is at least as large as that of any reward.}.  The Stoic 
considers that following any game there will be a branch in which
$\$max$ has tunneled into his pocket, and he then expects that, 
whatever game he might choose, he will receive $\$max$.  So his
preference order for games is:
\begin{quote}
{\bf Stoic preference order:} All games are preferred equally.\end{quote}
The Stoic acts as if all games have a non-zero value of $w(\$max)$.
If all games {\em really did} have a small-but-non-zero value of
$w(\$max)$, that would violate another of Wallace's axioms, the one called
``branching availability''.  Perhaps this is a good reason to question
that axiom also, but for the purposes of this note is is simpler to agree
that there {\em are} games for which  $w(\$max)$ is  precisely zero 
but that, due to the precision limitation, an agent can never know that he
is playing one of them.  Classically, it is also true that one can never 
be completely certain about the possible outcomes of a game; 
if you play a game in which
you are supposed to gain $\$1$, it might happen that the $\$1$-bill you
are given turns out to be an old and rare one which is worth $\$10^6$.
Usually one would judge that the chance of that happening is so small that
it need not be considered when deciding whether or not to play that game.
In the quantum case, the Born rule implies that branches with minuscule 
quantum weight can be neglected, but this cannot be assumed in an
argument meant to justify the Born rule\footnote{As noted by 
Price [3]}---and invoking the 
solution-continuity axiom does not
help, since the Stoic's preferences do satisfy that axiom.\footnote{I
imagine that Wallace's intention in adopting a continuity axiom is to 
force branches of exquisitely-small quantum weight to be neglected, 
which in the example above would mean roughly that the case of very
small $\epsilon$ must be treated the same as  the case in which 
$\epsilon$ is zero.  However, the Stoic turns this intention upside
down: he satisfies continuity by treating the case in which
$\epsilon$ is zero just as he does the case of very small $\epsilon$.}
In fact, these
preferences satisfy all of the axioms which Wallace has assumed in [2].
On the other hand, the Stoic acts in flagrant disregard for the DTBR.
The Born-rule expectations for $\$$ will be greater for game $A$
than for game $B_{\epsilon}$ defined above (and will remain greater even if 
a sufficiently-small contribution due to $\$max$ is included in the
calculation of those expectations) but the Stoic does not prefer playing
$A$ to playing $B_{\epsilon}$.  Even if he knows that the quantum weight for 
$\$max$ is, say, approximately $10^{-100}$, he acts as if that outcome had
probability one.

But what about  Wallace's proof?  The result that Wallace
does prove is given in eq.\ 30 of [2]; it says that there is 
an essentially unique function $u$ of rewards such that the agent prefers
a game $G$ to a game $H$ exactly when the quantum expectation of $u$
on $G$ is greater than the quantum expectation of $u$ on $H$.  This result 
is certainly correct for the Stoic: for him $u$ is a constant, the 
quantum expectation of a constant is the same for any game, and the
Stoic does not prefer any game to any other.

However, to get from this result to the DTBR requires the identification
of the function  $u$ with the agent's preference for rewards $\cal V$.
This might seem to be almost a matter of definition, since usually one
can establish an agent's preference for rewards from his preference
for games; for example,   if the agent does
not care whether he plays a game in which he surely receives a reward 
$r$ or a game
in which he surely receives a reward $s$, it usually follows that he 
does not care
whether he receives $r$ or $s$.  But for the Stoic in the MWI, this does
{\em not} follow; as I have defined him, he would prefer receiving $\$1$ 
to receiving $\$0$, 
but he does not prefer game $A$ to game $B_{0}$, because he expects
(rightly or wrongly) to receive $\$max$ from either game.

One can also imagine a classical situation where an agent's preference
for rewards does not follow from his preference for games:  Suppose that
an agent is offered a choice between two games, called $G$ and $H$; one
game will surely pay him $\$1$ and the other surely pay him $\$0$, but 
the agent does not know which of these games is the one which pays $\$1$.
Then, although he might prefer to receive the $\$1$, he would have no reason
to prefer the game called $G$ over the game called $H$.  That is, decision
theory applies in the usual way under the assumption that the agent
does know the rules of the games he is offered.  The 
Stoic's situation is analogous to this classical one;  because of the
precision limitation, he can
never know that he is being offered a game in which  $w(\$max)$
is precisely zero.  That is why his preference for rewards
does not follow from his preference for games. 

In order to justify the Born rule from the MWI, we might therefore want
to adopt some additional axiom in order to rule out the Stoic, which 
would mean adopting an axiom not because it was actually required
by rationality, but rather because it seemed to be required in order
to justify the Born rule.  This might be necessary in order to reach the
goal of identifying a set of assumptions which, when bundled together
with the MWI, would lead to the Born rule. However, it would not 
accomplish what I take to be the original goal of the D-W program,
which is to derive the Born rule from the MWI  with no
additional assumptions whatsoever.\footnote{Elsewhere [8] Wallace has
written ``The formalism is to be left alone\ldots unitary quantum
mechanics need not be supplemented in any way''.}

\vspace{0.8cm}
Acknowledgement: I would  like to acknowledge the hospitality of the
Lawrence Berkeley National Laboratory, where this work was done.

\end{document}